# Ultra-strong Adhesion of Graphene Membranes

Steven P. Koenig, Narasimha G. Boddeti, Martin L. Dunn, and J. Scott Bunch*

*Department of Mechanical Engineering, University of Colorado, Boulder, CO 80309 USA*

*email: jbunch@colorado.edu

**As mechanical structures enter the nanoscale regime, the influence of van der Waals forces increases. Graphene is attractive for nanomechanical systems[1,2] because its Young's modulus and strength are both intrinsically high, but the mechanical behavior of graphene is also strongly influenced by the van der Waals force[3,4]. For example, this force clamps graphene samples to substrates, and also holds together the individual graphene sheets in multilayer samples. Here we use a pressurized blister test to directly measure the adhesion energy of graphene sheets with a silicon oxide substrate. We find an adhesion energy of $0.45 \pm 0.02$ J/m$^2$ for monolayer graphene and $0.31 \pm 0.03$ J/m$^2$ for samples containing 2-5 graphene sheets. These values are larger than the adhesion energies measured in typical micromechanical structures and are comparable to solid/liquid adhesion energies[5-7]. We attribute this to the extreme flexibility of graphene, which allows it to conform to the topography of even the smoothest substrates, thus making its interaction with the substrate more liquid-like than solid-like.**

Figure 1a shows optical images of the devices used for this study. Graphene sealed microcavities were fabricated by the mechanical exfoliation of graphene over predefined wells (diameter ~5 um) etched in a SiO$_2$ substrate (See Methods). Two exfoliated graphene flakes were used, yielding membranes with between 1 and 5 graphene layers, which were suspended over the wells and clamped to the SiO$_2$ substrate by the van der Waals force.

After exfoliation the internal pressure in the microcavity, $p_{int}$, is equal to the external pressure, $p_{ext}$, which is atmospheric pressure. In this state the membrane is flat, adhered to the substrate, and it confines N gas molecules inside the microcavity.

To create a pressure difference across the graphene membrane, we put the sample in a pressure chamber and use nitrogen gas to increase $p_{ext}$ to $p_0$. Devices are left in the pressure chamber at $p_0$ for between 4 and 6 days in order for $p_{int}$ to equilibrate to $p_0$ (Fig. 1b). This is thought to take place through the slow diffusion of gas through the $SiO_2$ substrate[3]. We then remove the device from the pressure chamber, and the pressure difference ($p_{int} > p_{ext}$) causes the membrane to bulge upwards and the volume of the cavity to increase (Fig. 1c). We use an atomic force microscope (AFM) to measure the shape of the graphene membrane, which we parameterize by its maximum deflection, $\delta$, and its radius, $a$ (Fig. 1d).

This technique allows us to measure $\delta$ and $a$ for different values of $p_0$. Figure 1e shows a series of AFM line cuts through the center of a mono-layer membrane as $p_0$ is increased. At low $p_0$, the membrane is clamped to the substrate by the van der Waals force and $\delta$ increases with increasing $p_0$. At higher $p_0$ (e.g., $p_0 > 2$ MPa) in addition to an increased deflection, we also observe delamination of the graphene from the $SiO_2$ substrate which leads to an increase in $a$ (Fig. 1e). In Fig. 2a, we plot $\delta$ vs. $p_0$ for all the bilayer membranes measured; results are similar for other devices (see Supplementary Information). The deflection increases nonlinearly until $p_0 \sim 2.5$ MPa where $\delta$ then begins to increase more rapidly. The blister radius stays constant until $p_0 \sim 2.5$ MPa and then abruptly increases with increasing $p_0$ (Fig. 2b).

At large $p_0$ (e.g., $> 3.0$ MPa), stable delamination occurs: $a$ increases and thus $\Delta p$

decreases with increasing $p_0$ (Fig. 2c). All of the pressurized graphene membranes show a great degree of axisymmetry in their deformation before and after delamination. Stable delamination is in stark contrast to the common constant pressure blister test which results in unstable crack growth at the onset of delamination[8]. As a result we call this the *constant N blister test*, since the number of molecules in the microcavity is constant during blister delamination. While a macroscopic counterpart of the constant N blister test has been demonstrated[9], although not widely used, the novelty here is in the use of the adhesion between graphene and $SiO_2$ to prepare an impermeable seal for gas in the microcavity – filling and emptying of the microcavity are accomplished via diffusion through $SiO_2$ which is slow enough to allow reliable measurements of stable delamination[3].

We use the measured membrane profile ($\delta$ and $a$ vs. $p_0$) in the constant N blister test to determine the graphene/$SiO_2$ adhesion energy. To this end, we describe the deformation of the membrane using Hencky's solution[10,11] for the geometrically nonlinear response of a clamped isotropic circular elastic membrane subjected to a pressure difference $\Delta p$ across the membrane. This solution provides the membrane profile in the form of an infinite series in radial position, as well as the relationship between the pressure difference and blister height, $\Delta p = K(\upsilon)\frac{\delta^3}{a^4}Et$, and the volume of the blister $V_b(a) = C(\upsilon)\pi a^2\delta$. Here $E$ is Young's modulus, $\upsilon$ is Poisson's ratio, $t$ is the membrane thickness, and $C(\upsilon)$ and $K(\upsilon)$ are coefficients that only depend on $\upsilon$ and vary from $K(\upsilon = 0.10) = 2.93$ to $K(\upsilon = 0.20) = 3.22$. The $K(\upsilon)\frac{\delta^3}{a^4}$ term primarily describes the geometrical nonlinear deflection-pressure response of the circular membrane as $K(\upsilon)$ is a coefficient that is fixed for a specified $\upsilon$. For graphene, we take $\upsilon = 0.16$[12] and so $K(\upsilon = 0.16) = 3.09$ and $C(\upsilon = 0.16) = 0.524$.

To determine the adhesion energy we model the constant N blister as a thermodynamic system with free energy:

$$F = \frac{(p_{int} - p_{ext})V_b}{4} + \Gamma\pi(a^2 - a_o^2) - p_oV_o\,ln[\frac{V_o + V_b}{V_o}] + p_{ext}V_b \qquad (1)$$

where $V_o$ is the initial volume of the microcavity, $\Gamma$ is the graphene/SiO$_2$ adhesion energy, and $a_0$ is the initial radius before delamination[9]. In eq. (1) the four terms represent, respectively, i) stretching of the membrane due to the pressure difference across it, $\Delta p = p_{int} - p_{ext}$; we calculate it by equating the strain energy in the deformed membrane to the work done by the expanding gas during deformation (which is easier to directly calculate) and then simplifying the results using Hencky's relations for the pressure-deflection and pressure-blister volume; ii) graphene/SiO$_2$ adhesion; iii) expansion of the gas in the chamber from $V_o$ to a final volume $V_o + V_b(a)$; and iv) work done on the gas held at a fixed external pressure $p_{ext}$. To deduce $\Delta p$ across the membrane we use the ideal gas law and assume isothermal expansion of the trapped gas (see Methods). Minimizing the free energy with respect to $a$, provides a relationship between $\Gamma$, $\delta$, and $a$:

$$\Gamma = \frac{5c}{4}(p_o\frac{V_o}{V_o + V_b(a)} - p_{ext})\delta \qquad (2)$$

We use eq. (2) to determine $\Gamma$ with prescribed values of p$_o$ and p$_{ext}$, $(a, \delta)$ pairs measured by AFM, V$_o$ determined by the cavity geometry, and $V_b(a)$. Values of adhesion energy extracted in this manner for all devices are shown in Fig. 3. The value $\Gamma = 0.31 \pm 0.03$ J/m$^2$ describes the multilayer graphene/SiO$_2$ adhesion reasonably well for both SiO$_2$ substrates used in this study, but not the monolayer which has a value of $0.45 \pm 0.02$ J/m$^2$ (see Supplementary Information).

Our measured adhesion energies are approximately four orders of magnitude larger than adhesion energies commonly found in micromechanical systems where van der Waals forces across noncontacting regions between asperities play a significant role and approximately five times larger than adhesion in gold coated submicron beams[5,7,13-15]. They are also twice that of previous estimates for multilayer graphene to a $SiO_2$ substrate[16], however, those results are extracted from a model that uses an estimate of Young's modulus of graphene that is one-half of that measured here. Our results are comparable to values deduced from experiments on collapsed carbon nanotubes[17]. Using values derived from the measured surface energies of graphite ($\gamma = 165\text{-}200$ mJ/m$^2$) and $SiO_2$ ($\gamma = 115\text{-}200$ mJ/m$^2$), one expects an adhesion energy of $\Gamma = 2\ (\gamma_{SiO2}\ \mathrm{x}\ \gamma_{graphite})^{1/2} = (0.275 - 0.4)$ J/m$^2$ [6,17]. The close agreement between our measured adhesion energy and this estimate suggests that graphene makes close and intimate contact with the $SiO_2$ substrate[18,19]. It shows that atomically thin structures like graphene demonstrate conformation over the $SiO_2$ surface that is more reminiscent of a liquid than a solid.

The reason for the higher adhesion of monolayer graphene to multilayer graphene is not entirely understood. We ruled out bonding due to induced image charges from buried charges in the $SiO_2$ substrate (see Supplementary Information). A possible explanation for the discrepancy between 1 and 2-5 layers is the increased ability of monolayer graphene to conform to the contours of the surface due to its flexibility. Roughness measurements of various layers of graphene on the $SiO_2$ substrate taken with the AFM show a decreasing roughness with increasing layer number (about 197 pm for bare $SiO_2$, 185 pm with one layer, and 127 pm with 15 layers of graphene) suggesting that monolayer graphene conforms more closely to the $SiO_2$ substrate (see Supplementary Information). Recent theory that idealizes the substrate roughness as a sinusoidal profile shows a jump in adhesion energy with wavelength and amplitude[20-22]. We modified this theory to account for effects of multilayer graphene and it supports the suggestion of a jump to contact that results in increased adhesion energy as the number of layers decreases, however the model is too simple to quantitatively predict that this jump occurs between N = 2 and 1 layers.

As mentioned, the deformation of the membrane can be described using Hencky's solution for the geometrically nonlinear response of a clamped circular elastic membrane subjected to a pressure difference $\Delta p$ across the membrane. The dashed line in Fig. 1e compares the calculated profile using Hencky's solution[10,11] with our measured profile. The close agreement validates the use of $a$ and $\delta$ to parameterize the deformation. Figure 2c shows the equilibrium $p_{int}$ vs. $p_o$ for the bilayer devices. The solid lines in Figure 2a and 2c are the solutions of $\Delta p = K(v)\frac{\delta^3}{a^4}Et$ for a constant $a = a_0$ (no delamination) where we used the fitted value of $Et$. This provides a good fit until delamination begins ($a > a_0$) at $p_o$ = 2.5 MPa (Fig. 2b). The dashed lines in Fig. 2 are theoretical predictions of $\delta$, $a$, and $p_{int}$ vs $p_o$ using the average adhesion energy values from Fig. 3 and the fitted value of $Et$.

Figure 4a shows $\Delta p$ vs $K(v)\frac{\delta^3}{a^4}$ for the monolayer graphene membrane as well as a linear fit to eq. (3) to determine $Et$ = 347 N/m. This agrees well with previous measurements for graphene and the in plane modulus ($E$ = 1 TPa) and interatomic spacing of graphite ($t$ = 0.335 nm) [3,4,12]. Figure 4b-4e shows $\Delta p$ vs $K(v)\frac{\delta^3}{a^4}$ for multilayer membranes. Included are linear fits to the data for $\Delta p < 0.50$ MPa (dashed lines). Theoretical estimates with $nEt$ (solid lines) where $Et$ = 347 N/m (our monolayer measurement) and $n$ = 1-5 corresponds to the number of graphene layers are also plotted and the $Et$ values obtained by both methods are compared in Fig. 4f. The good agreement between these values demonstrate that the additional graphene layers are sufficiently well-adhered to the substrate and each other by the van der Waals force so that the pressure load is carried by all the layers and no significant sliding or delamination occurs up to pressures as large as $\Delta p$ = 0.50 MPa[23,24]. For $\Delta p$ < 0.25 MPa the effect of initial tension in the membrane cannot be neglected and for $\Delta p$ > 0.50 MPa the data shows considerably more scatter (see Supplementary Information). Further work is necessary to understand the origin of this scatter, but two possibilities are small amounts of sliding or early stages of delamination which are difficult to measure by AFM.

In conclusion, we demonstrated a simple yet reliable constant N blister test and used it to measure the adhesion energy of the thinnest nanostructures possible, single and multilayer graphene sheets, to $SiO_2$. This is the first direct measurement of the adhesion energy of 1-5 layer graphene to $SiO_2$ – a substrate on which the majority of graphene electrical and mechanical devices are fabricated. This result can be used to guide developments in graphene based electrical and mechanical devices where adhesive forces are known to play an important role as well as provide opportunities for fundamental studies of surface forces in the thinnest structures possible[3,25-28].

### Methods

Suspended graphene membranes are fabricated by a combination of standard photolithography and mechanical exfoliation of graphene. First, an array of circles with diameters of 5 μm and 7 μm are defined by photolithography on an oxidized silicon wafer with a silicon oxide thickness of 285 nm. Reactive ion etching is then used to etch the circles into cylindrical cavities with a depth of 250-300 nm leaving a series of microcavities on the wafer. Mechanical exfoliation of natural graphite using Scotch tape is then used to deposit suspended graphene sheets over the microcavities[29]. Of the 39 membranes there were 5 1-layer, 10 2-layer, 15 3-layer, 4 4-layer, and 5 5-layer membranes. The number of graphene layers was verified using a combination of Raman spectroscopy, optical contrast, AFM measurements, and elastic constants measurements (see Supplementary Information)[30,31]. Two flakes on two different $SiO_2$ substrates were used in this study (Fig. 1a). Three 2-layer membranes, four 3-layer membranes and one 4 layer membrane were damaged before reaching the highest pressures.

To deduce $\Delta p = p_{int} - p_{ext}$ across the membrane we use the ideal gas law and assume isothermal expansion of the trapped gas with a constant number of molecules, N. Doing so leads to $p_o V_o = p_{int}(V_o + V_b)$ where $V_o$ is the initial volume of the microcavity and $V_b$ is the volume of the pressurized blister after the device is brought to atmospheric pressure and bulges upward. The assumption of constant N is valid considering that the deflection does not change over the ~ 20 minutes that the AFM images are acquired suggesting that no significant change in N, due to gas "leaking", occurs at the time scale of the experiment.

**References:**

1.    Bunch, J.S. et al. Electromechanical Resonators from Graphene Sheets. *Science* **315**, 490-493 (2007).

2.    Meyer, J.C. et al. The structure of suspended graphene sheets. *Nature* **446**, 60-63 (2007).

3.    Bunch, J.S. et al. Impermeable atomic membranes from graphene sheets. *Nano Lett.* **8**, 2458-2462 (2008).

4.  Lee, C. et al. Measurement of the Elastic Properties and Intrinsic Strength of Monolayer Graphene. *Science* **321**, 385-388 (2008).

5.  Maboudian, R. & Howe, R.T. Critical Review: Adhesion in surface micromechanical structures. *J. Vac. Sci. Tech. B* **15**, 1-20 (1997).

6.  Israelachvilli, J. *Intermolecular and Surface Forces*. (Academic Press: 2011).

7.  Delrio, F.W. et al. The role of van der Waals forces in adhesion of micromachined surfaces. *Nature Mater.* **4**, 629-634(2005).

8.  Gent, A.N. & Lewandowski, L.H. Blow-off pressures for adhering layers. *J. of Appl. Polym. Sci.* **33**, 1567-1577 (1987).

9.  Wan, K. & Mai, Y. Fracture mechanics of a new blister test with stable crack growth. *Acta Metall. Mater.* **43**, 4109-4115 (1995).

10. Hencky, H. Über den spannungszustand in kreisrunden platten mit verschwindender biegungssteiflgkeit. *Zeitschrift fur Mathematik und Physik* **63**, 311-317 (1915).

11. Williams, J. Energy release rates for the peeling of flexible membranes and the analysis of blister tests. *Int. J. Fracture* **87**, 265-288 (1997).

12. Blakslee, O.L. et al. Elastic Constants of Compression-Annealed Pyrolytic Graphite. *J. Appl. Phys.* **41**, 3373-3382 (1970).

13. DelRio, F.W. et al. The effect of nanoparticles on rough surface adhesion. *J. Appl. Phys.* **99**, 104304 (2006).

14. DelRio, F.W. et al. Elastic and adhesive properties of alkanethiol self-assembled monolayers on gold. *Appl. Phys. Lett.* **94**, 131909 (2009).

15. Buks, E. & Roukes, M.L. Stiction, adhesion energy, and the casimir effect in micromechanical systems. *Phys. Rev. B* **63**, 33402 (2001).

16. Zong, Z. et al. Direct measurement of graphene adhesion on silicon surface by intercalation of nanoparticles. *J. Appl. Phys.* **107**, 026104 (2010).

17. Yu, M.F., Kowalewski, T. & Ruoff, R.S. Structural analysis of collapsed, and twisted and collapsed, multiwalled carbon nanotubes by atomic force microscopy. *Phys. Rev. Lett.* **86**, 87-90 (2001).

18. Cullen, W. et al. High-Fidelity Conformation of Graphene to $SiO_2$ Topographic Features. *Phys. Rev. Lett.* **105**, 215504 (2010).

19. Rudenko, A.N. et al. Local interfacial interactions between amorphous $SiO_2$ and supported graphene. *arXiv:1105.1655v1* (2011).

20. Aitken, Z.H. & Huang, R. Effects of mismatch strain and substrate surface corrugation on morphology of supported monolayer graphene. *J. Appl. Phys.* **107**, 123531 (2010).

21. Li, T. & Zhang, Z. Substrate-regulated morphology of graphene. *J. Phys. D Appl. Phys.* **43**, 075303 (2010).

22. Viola Kusminskiy, S. et al. Pinning of a two-dimensional membrane on top of a patterned substrate: The case of graphene. *Phys. Rev. B* **83**, 165405 (2011).

23. Suk, J.W. et al. Mechanical Properties of Monolayer Graphene Oxide. *ACS Nano* **4**, 6557-6564 (2010).

24. Ruiz-Vargas, C.S. et al. Softened Elastic Response and Unzipping in Chemical Vapor Deposition Graphene Membranes. *Nano Lett.* **11** (6), 2259-2263 (2011).

25. Lee, C. et al. Frictional characteristics of atomically thin sheets. *Science* **328**, 76-80(2010).

26. Lui, C.H. et al. Ultraflat graphene. *Nature* **462**, 339-41 (2009).

27. Capasso, F. et al. Casimir forces and quantum electrodynamical torques: Physics and nanomechanics. *IEEE J. Sel. Top. Quant.* **13**, 400-414 (2007).

28. Lu, Z. & Dunn, M.L. van der Waals adhesion of graphene membranes. *J. Appl. Phys.* **107**, 044301 (2010).

29. Novoselov, K.S. et al. Two-dimensional atomic crystals. *P. Natl. Acad. Sci. USA* **102**, 10451-10453 (2005).

30. Koh, Y.K. et al. Reliably Counting Atomic Planes of Few-Layer Graphene (n > 4). *ACS Nano* **5**, 269-274 (2011).

31. Ferrari, A.C. et al. Raman Spectrum of Graphene and Graphene Layers. *Phys. Rev. Lett.* **97**, 187401 (2006).

Acknowledgements:

This work was supported by the National Science Foundation (NSF: Grant numbers 0900832 and 1054406), the NSF Industry/University Cooperative Research Center for Membrane Science, Engineering and Technology at the University of Colorado at Boulder, and the DARPA Center on Nanoscale Science and Technology for Integrated Micro/Nano- Electromechanical Transducers (DARPA/SPAWAR; grant number N66001-10-1-4007). Sample fabrication was performed at the University of Colorado node of the National Nanofabrication Users Network, funded by the NSF. We thank Guillermo Acosta, Luda Wang and Xinghui Liu for help with fabrication and Rishi Raj for use of the Raman microscope.

**Author Contributions:** S.P.K. performed the experiments. S.P.K. and J.S.B. conceived and designed the experiments. N.G.B. and M.L.D. developed the theory and modelling. All authors interpreted the results and co-wrote the manuscript.

**Additional Information:**

Supplementary information accompanies this paper at www.nature.com/naturenanotechnology. Reprints and permission information is available online at http://npg.nature.com/reprintsandpermissions/. Correspondence and requests for materials should be addressed to J.S.B.

**Figure Captions**

Figure 1. **Pressurizing Graphene Membranes**

(a) Two Optical images showing graphene flakes with regions of 2-5 suspended layers (top) and 1 and 3 suspended layers (bottom). The arrays of microcavities in the $SiO_2$ substrate can also be seen. The number of graphene layers was verified with a combination of Raman spectroscopy, atomic force microscopy and measurements of optical contrast and elastic constants measurements (see supplementary information).

(b) Schematic illustration of a graphene-sealed microcavity before it is placed in the pressure chamber. The pressure inside the microcavity $p_{int}$ is equal to the external pressure $p_{ext}$, so the membrane is flat. After 4-6 days inside the pressure chamber, $p_{int}$ increases to $p_0$.

(c) When the microcavity is removed from the pressure chamber, the pressure difference across the membrane causes it to bulge upward and eventually delaminate from the substrate, causing the radius $a$ to increase.

(d) Three dimensional rendering of an AFM image showing the deformed shape of a monolayer graphene membrane with $\Delta p = p_{int} - p_{ext} = 1.25$ MPa.

(e) Deflection versus position as $\Delta p$ is increased from 0.145 MPa (black) to 1.25 MPa (cyan). The dashed black line is the shape obtained from Hencky's solution for $\Delta p$ =0.41 MPa. The deflection is measured by an AFM along a line that passes through the centre of the membrane.

Figure 2 **Delaminating Graphene Membranes**

(a-c) Plots showing the maximum deflection $\delta$ (a), the blister radius $a$ (b) and the internal pressure $p_{int}$ (c) versus the pressure inside the pressure chamber $p_0$ for all the two-layer membranes we studied. The solid black line is a theoretical curve assuming no delamination of the membrane. The dashed lines are the calculated theoretical curves for $nEt$ = 694 N/m where $n$ = 2 and adhesion energies of 0.25 J/m$^2$, 0.31 J/m$^2$ and 0.37 J/m$^2$.

Figure 3 **Measured Graphene/SiO$_2$ Adhesion Energies**

Measured adhesion energies $\Gamma$ for graphene membranes containing 1 layer (black circles), 2 layers (red squares), 3 layers (green triangles), 4 layers (blue triangles) and 5 layers (cyan diamonds). The upper solid line corresponds to $\Gamma$ = 0.45 J/m$^2$ and the lower dashed line corresponds to $\Gamma$ = 0.31 J/m$^2$.

Figure 4 **Elastic Constants and Clamping of Graphene Membranes**

(a-e) K($v$)($\delta^3/a^4$) versus pressure difference $\Delta p$ for membranes containing 1-5 graphene sheets before delamination (using the same colour scheme as figure 3) and after delamination (magenta symbols in all plots). The solid lines are linear fits to all the data with $nEt$ = 347 N/m (black), 694 (red), 1041 (green), 1388 (blue) and 1735 N/m (cyan). The dashed lines show linear fits to the data for $\Delta p$ < 0.50 MPa and have slopes corresponding to $Et$ = 661 (red; 2 layers), 950 (green; 3 layers), 1330 (red; 4 layers) and 1690 N/m (cyan; 5 layers). Note that the vertical scales are different

(f) $nEt$ versus number of layers. Closed shapes are for the fitted lines; open shapes are for $nEt$ where n is the number of layers and $Et$ = 347.

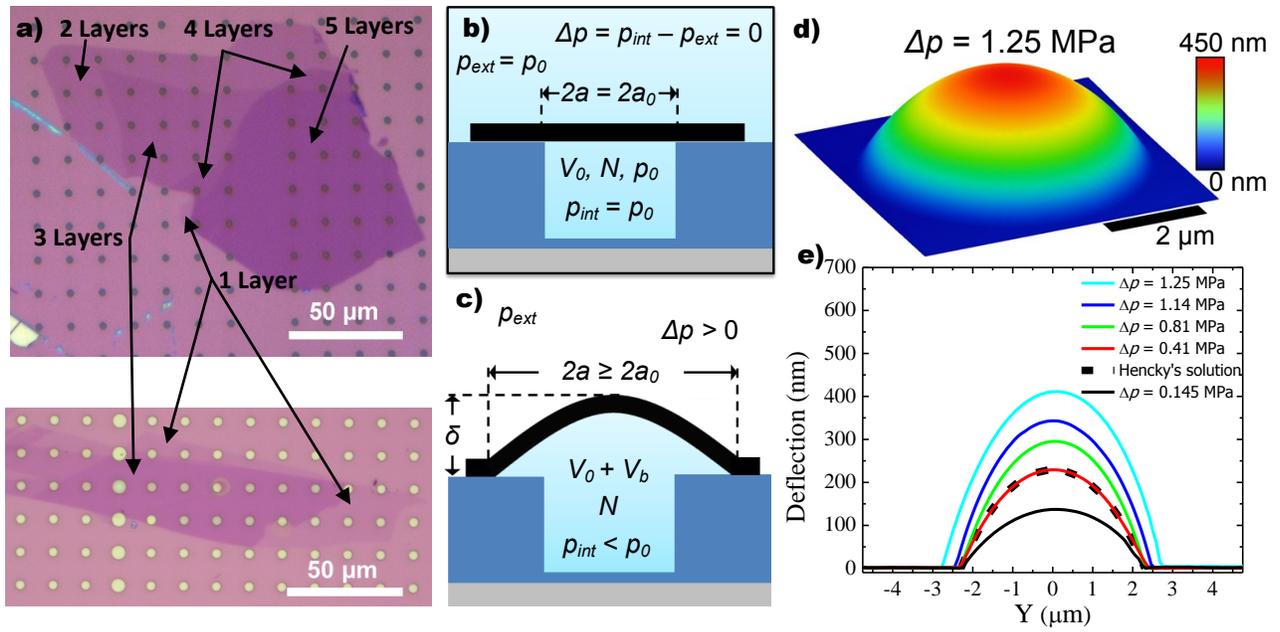

**Figure 1**

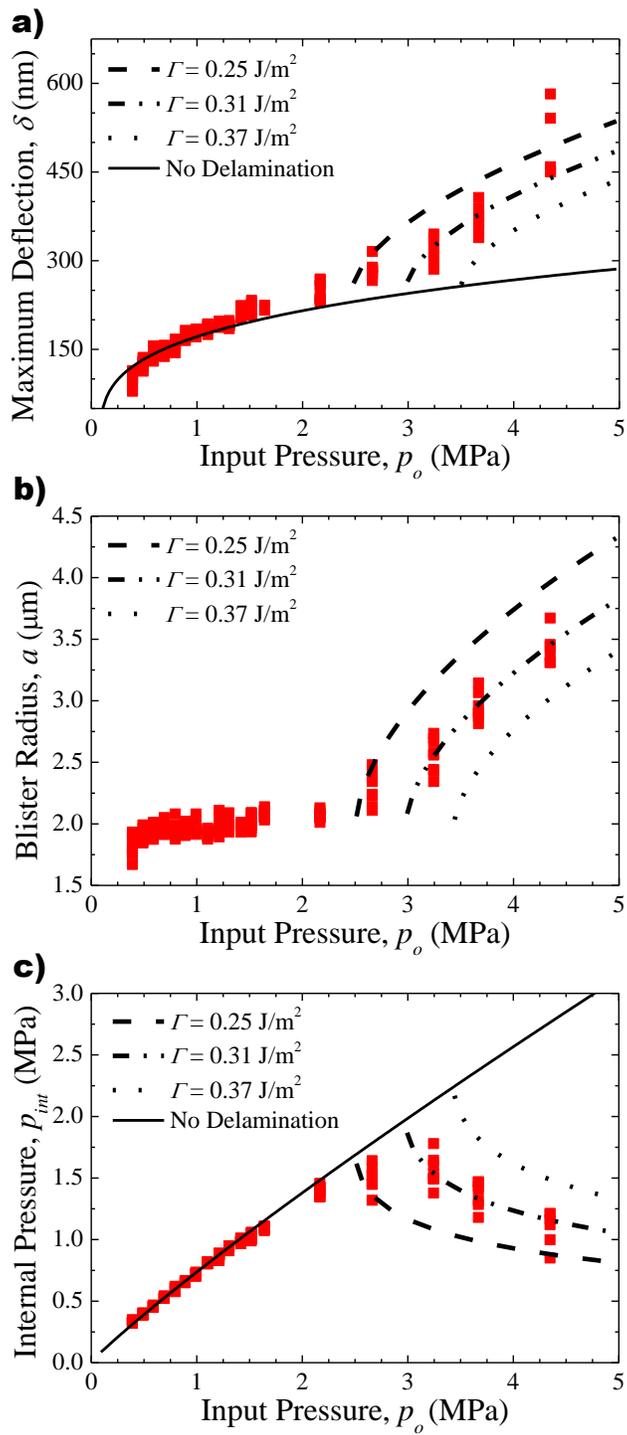

**Figure 2**

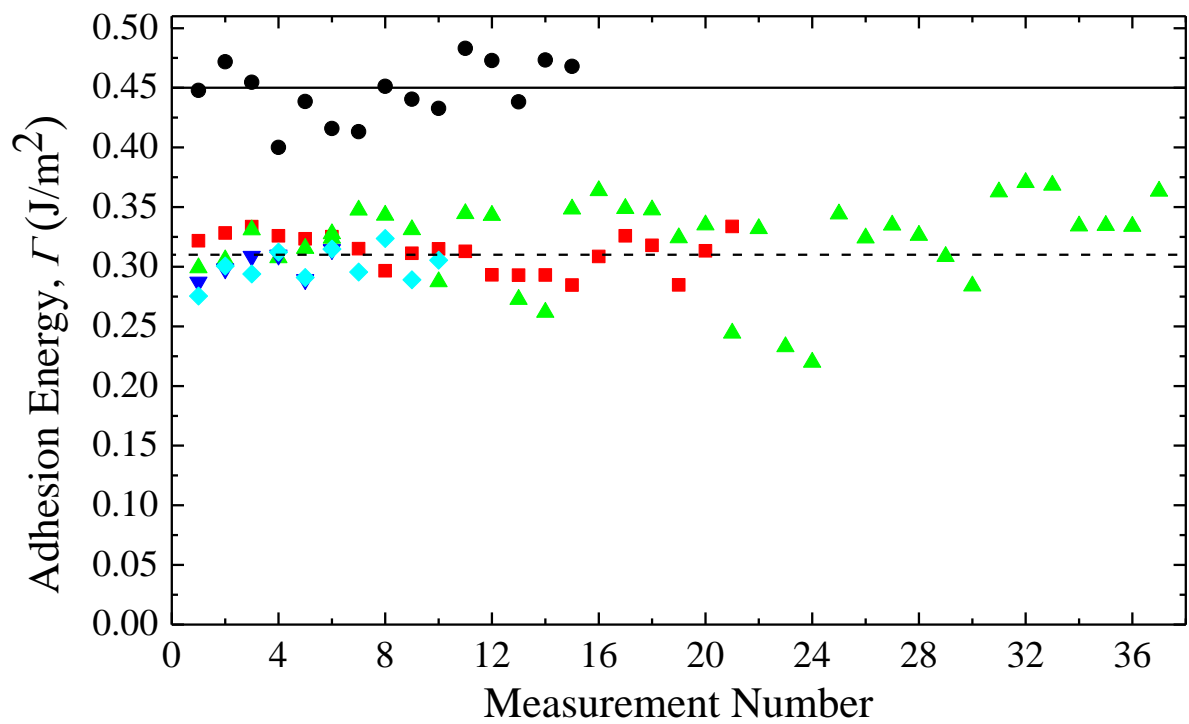

**Figure 3**

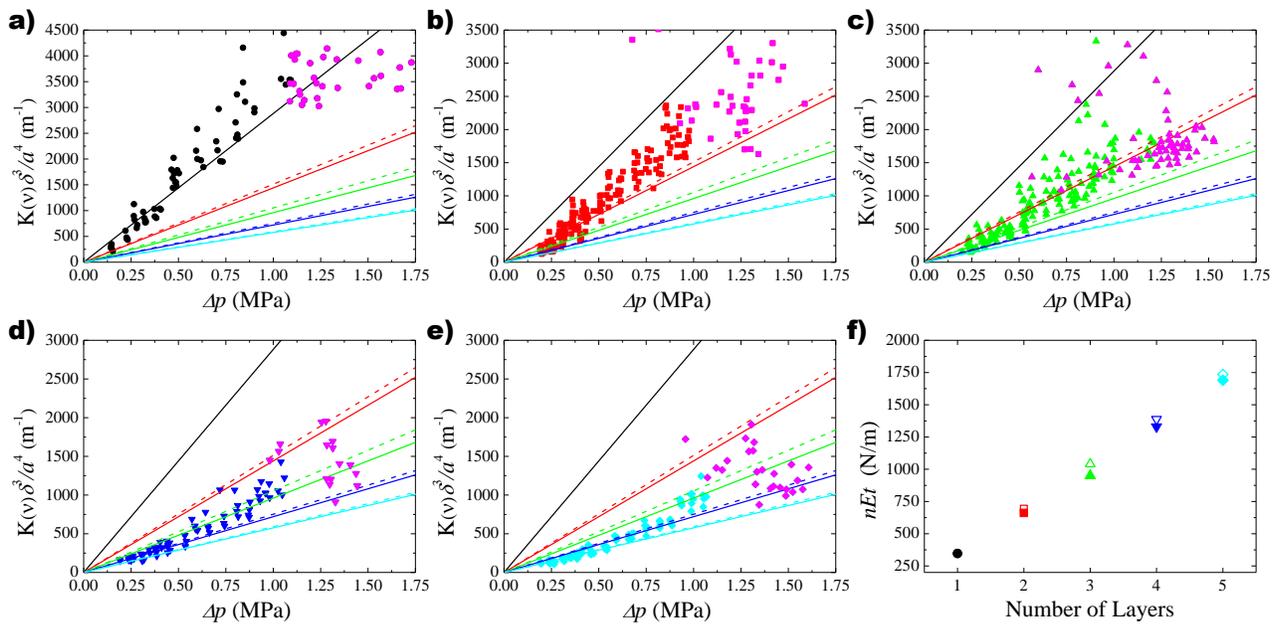

**Figure 4**

# Ultra-strong Adhesion of Graphene Membranes

Steven P. Koenig, Narasimha G. Boddeti, Martin L. Dunn, and J. Scott Bunch*

*Department of Mechanical Engineering, University of Colorado, Boulder, CO 80309 USA*

*email: jbunch@colorado.edu

**Supplementary Information:**

### Counting Number of Graphene Layers

In order to count the number of graphene layers used in this study we employed a combination of optical contrast, Raman spectroscopy, AFM measurements, and the elastic constant measurements. Raman spectroscopy has been demonstrated to be a powerful tool for identifying single layer graphene sheets [1]. Recently Raman has also been shown to be able to identify the number of layers of few layer graphene, a technique we use here[2]. Figure S1 (a) and (b) show the graphene flakes from this study and the spots where Raman spectrum was taken for each device, black is 1 layer, red is 2 layers, green is 3 layers, blue is 4 layers and cyan is 5 layers. Figure S1 (c) and (d) show the Raman spectrum taken from the spots of corresponding color in (a) and (b) respectively. To verify the number of layers we found the ratio of the integrated intensity of the first order optical phonon peak and the graphene G peak. The ratios are shown in figure S1 (e) and (f). Comparing these values with the Fresnel equation we can determine the number of layers for each region. In order to verify this technique we used optical contrast, AFM measurements, as well as the elastic constants of the membranes [3]. The optical contrast and AFM measurements showed close agreement to the Raman spectroscopy technique validating its utility.

**Adhesion Energy and Elastic Constants Measurements**

The adhesion energy measurements were carried out according to the main text of this article. Figures S2, S3, and S4 show (a) $\delta$ vs. $p_0$, (b) $a$ vs. $p_0$, and (c) $p_{int}$ vs. $p_0$, for all the membranes studied. The layer numbers are as follows: (a) 1 layer membranes from Fig. 1 (lower). (b) 2 layer membranes from Fig. 1 (upper). (c) 3 layer membranes from Fig. 1 (upper). 4 layer membranes from Fig. 1 (upper). (d) 5 layer membranes from Fig. 1 (upper). and (e) 3 layer membranes from Fig. 1 (lower).

**Repeatability of Elastic Constant Measurements**

To verify the repeatability of the measurement of the elastic constants at $\Delta p < 0.5$ MPa we first pressurized the graphene flake in Fig. 1a(upper) up to $\Delta p = 0.45$ MPa and then let pressure decrease back to $\Delta p = 0$ MPa. We then repeated the measurements and increased $\Delta p$ until there was significant peeling from the substrate in order to test the adhesion strength. Figure S5 shows the results from this test for (a) 2 layers, (b) 3 layers, (c) 4 layers, and (d) 5 layers of graphene. From this we conclude that pressurizing the membranes does not cause sliding or change the membrane properties when $\Delta p < 0.5$ MPa and therefore the membrane can be considered to be well clamped to the substrate in this pressure range.

**Adhesion from Trapped Charges in SiO$_2$**

We use the method of image charges to estimate the influence of trapped charges in the SiO$_2$ on the adhesion of graphene to the substrate. The work needed to move a charge from a distance $d$ from the conducting plane out to infinity is:

$$W = \frac{1}{4\pi\varepsilon_o}\frac{q^2}{4d} \quad \text{(S1)}$$

where $q$ is the fundamental charge, $d$ is the distance the charge is away from the conducting plane and $\varepsilon_o$ is the permittivity of free space[4]. In order to determine an adhesion energy we also need to know the area density of charges, $\rho$, and the equation becomes:

$$\Gamma = \frac{1}{4\pi\varepsilon_o}\frac{q^2}{4d}\rho \quad \text{(S2)}$$

If we assume all the charges are on the surface of the $SiO_2$ and that the equilibrium spacing between the graphene and $SiO_2$ is equal to that of the equilibrium spacing of graphite $d = 0.34$ nm. The charge density needed to produce our measured adhesion energy of 0.31 J/m$^2$ is ~9x10$^{17}$ m$^{-2}$. The charge density of $SiO_2$ is reported to be 2.3x10$^{15}$ m$^{-2}$ [5]. Seeing that the reported value of the charge density in $SiO_2$ is almost three orders of magnitude lower, we can conclude that trapped charges do not have a significant contribution to the adhesion energy value we measure. Other studies have used potassium ions to increase the charge density present in the oxide [6]. The concentration of potassium ions was as high as ~5 x 10$^{16}$ m$^2$. This upper limit of the extrinsic doping concentration results in a charge density that is one order of magnitude less than that needed to have adhesion energies on the order of what we measured. These results show that the effect of charge impurities in the $SiO_2$ below the graphene will not significantly influence our measure of adhesion energy.

**RMS Roughness and Conformation**

Roughness measurements were taken using a Veeco Dimension 3100 operating under non-contact mode under ambient conditions. The bare $SiO_2$ substrate is denoted as 0 layers in Fig. S6 and a ~5nm thick flake as measured by the AFM was estimated to be approximately 15 layers thick. For the roughness measurements of the substrate and each layer thickness multiple images were taken at various locations of each region, the images were taken from the chip in Fig. 1a (lower) and the RMS roughness was analysed using Wsxm software for each image [7]. The 1-3 layers were taken from the flake in Fig. 1a while the substrate measurements were taken from areas around the flake and the ~15 layer measurement was taken from a thick flake near the flake seen in Fig. 1a(lower). For the substrate and each different layer thickness, 7 images were used for the substrate, 4 images were used for 1 layer, 5 images for 2 layer, 3 images for 3 layers, and 2 images for the ~15 layer sample. Figure S6 shows the average roughness for the substrate, 0 layers, 1 layer, 2 layers 3 layers and ~15 layers as well as the standard deviation of the measurements shown by the error bars. These measurements suggest that graphene conforms more intimately to the substrate and as the number of layers is  decreased

**References:**

1. Ferrari, A.C. et al. Raman Spectrum of Graphene and Graphene Layers. *Phys. Rev. Lett.* **97**, 187401 (2006).

2. Koh, Y.K. et al. Reliably Counting Atomic Planes of Few-Layer Graphene (n > 4). *ACS Nano* **5**, 269-274 (2011).

3. Nair, R.R. et al. Fine structure constant defines visual transparency of graphene. *Science* **320**, 1308 (2008).

4. David J. Griffiths  *Introduction to Electrodynamics*. (Addison-Wesley: Upper Saddle River, 1999).

5. Martin, J. et al. Observation of electron–hole puddles in graphene using a scanning single-electron transistor. *Nature Phys.* **4**, 144-148 (2007).

6. Chen, J.-H. et al. Diffusive charge transport in graphene on $SiO_2$. *Solid State Commun.* **149**, 1080-1086 (2009).

7.    Horcas, I. et al. WSXM: A software for scanning probe microscopy and a tool for nanotechnology. *Rev. Sci. Instrum.* **78**, 13705 (2007).

**Supplementary Information Figures**

Figure S1. **Counting the Number of Layers**

(a) and (b) Optical images showing the graphene flakes used in this study. The colored circles denote the location at which Raman spectroscopy was taken (denoted as follows: black 1 layer, red 2 layers, green 3 layers, blue 4 layers, and cyan 5 layers)

(c) and (d) Raman spectrum from the graphene flakes in (a) and (b). The color of each curve corresponds to the spot on the optical image.

(e) and (d) Ratio of the integrated intensity of the first order silicon peak I(Si) and graphene G peak, I(G) (i.e. I(G)/I(Si)).

Figure S2. **Measured Deflection vs. Input Pressure**

(a) $-$ (f) $\delta$, vs $p_o$, for 1-5 layer devices. 1 layer devices (a) are from graphene flake in Fig. 1a(lower) and the 2-5, (b)-(e) respectively, are from the flake in Fig. 1a. (f) The data in f was determined to be 3 layers thick and taken from the lower graphene flake in Fig 1a.

Figure S3. **Blister Radius vs. Input Pressure**

(a) $-$ (f) $a$ vs. $p_o$ for 1-5 layer devices in Fig. S2.

Figure S4. **Internal Pressure vs. Input Pressure**

(a) $-$ (f) $p_{int}$ vs $p_o$ for 1-5 layer devices in Fig. S2.

Figure S5. **Repeatability of Measurements at Low Pressure Differences**

(a) - (d) K($\delta^3/a^4$) vs $\Delta p$ for 2-5 layer devices. The black points are from the first pressure cycling of the upper device in Fig. 1(a). After the highest pressure was measured the pressure was allowed to decrease back to atmospheric pressure and the measurements were repeated and carried higher pressures. This shows that up to $\Delta p \approx 0.5$ MPa there is no altering of the membrane properties between measurements.

Figure S6. **Measured Roughness of the Substrate**

RMS roughness measurements taken by non-contact AFM of the substrate (0 layers), 1, 2, and 3 layers as well a thick graphene sample that was ~5 nm (~15 layers) thick  as determined by the AFM. Error bars are ±1 standard deviation.

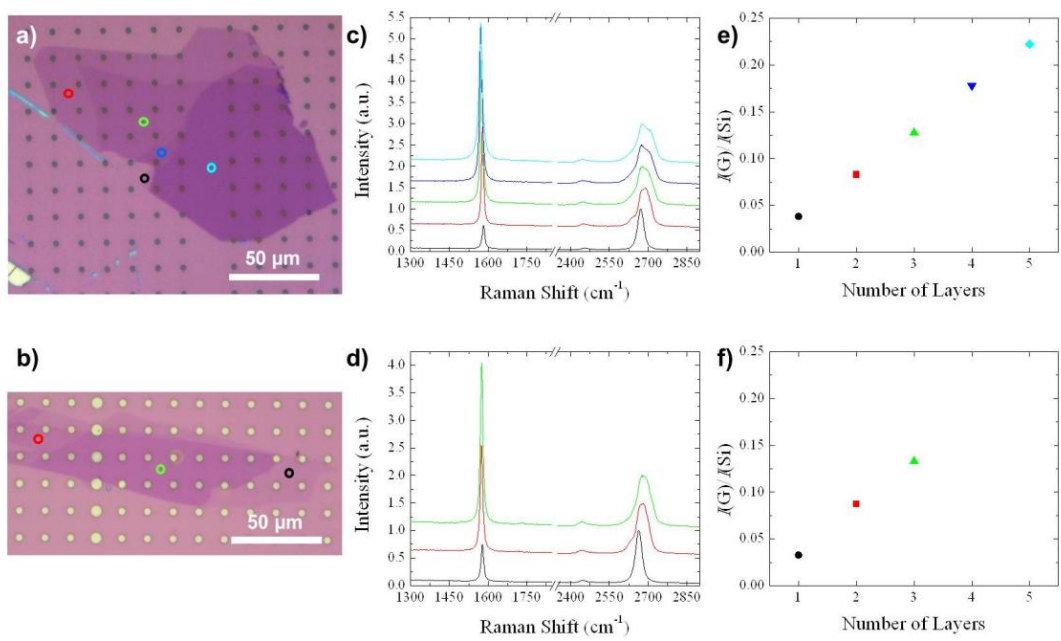

**Figure S1**

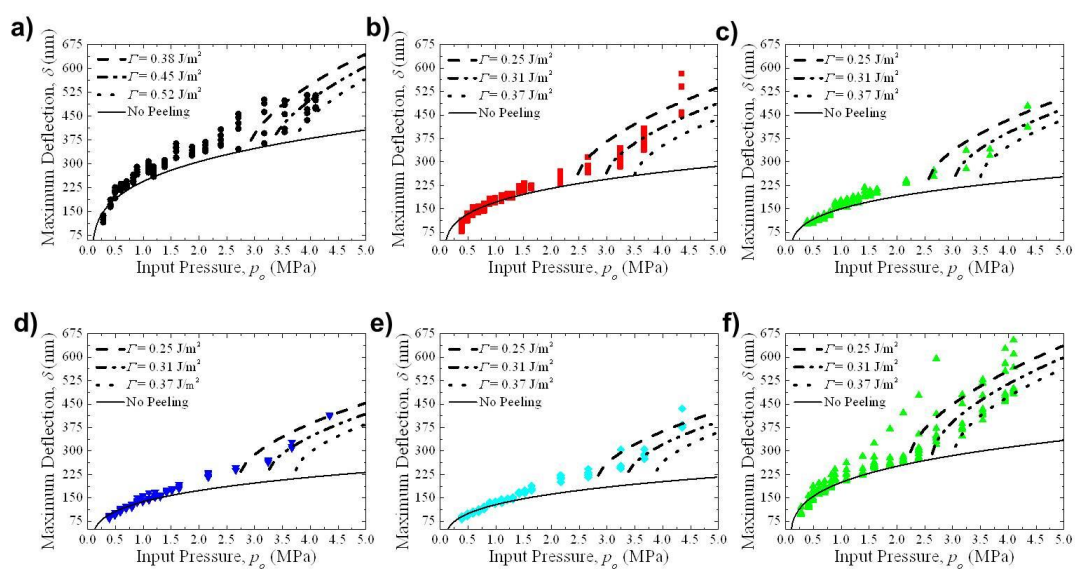

**Figure S2**

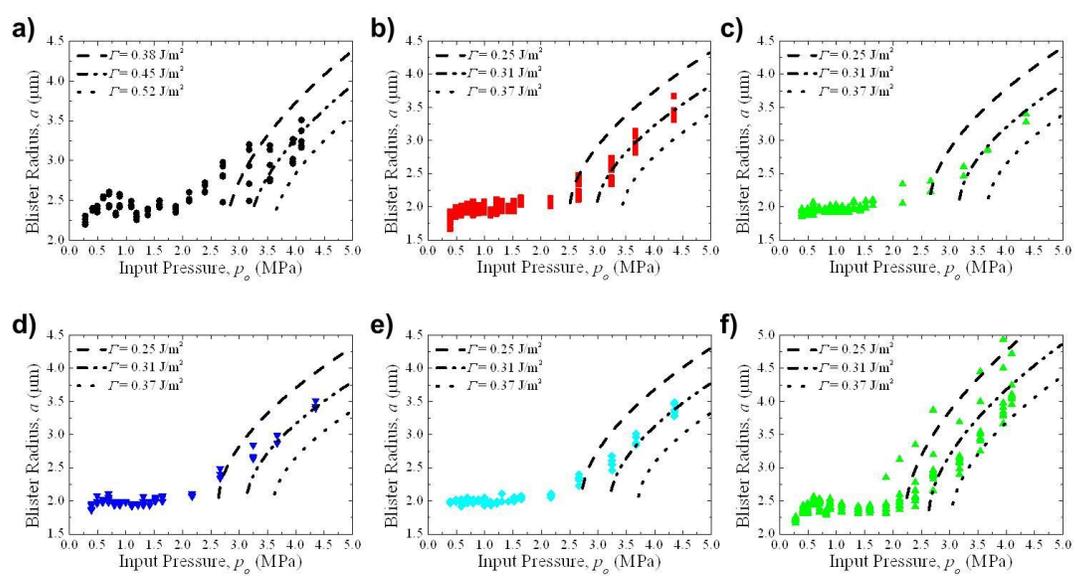

**Figure S3**

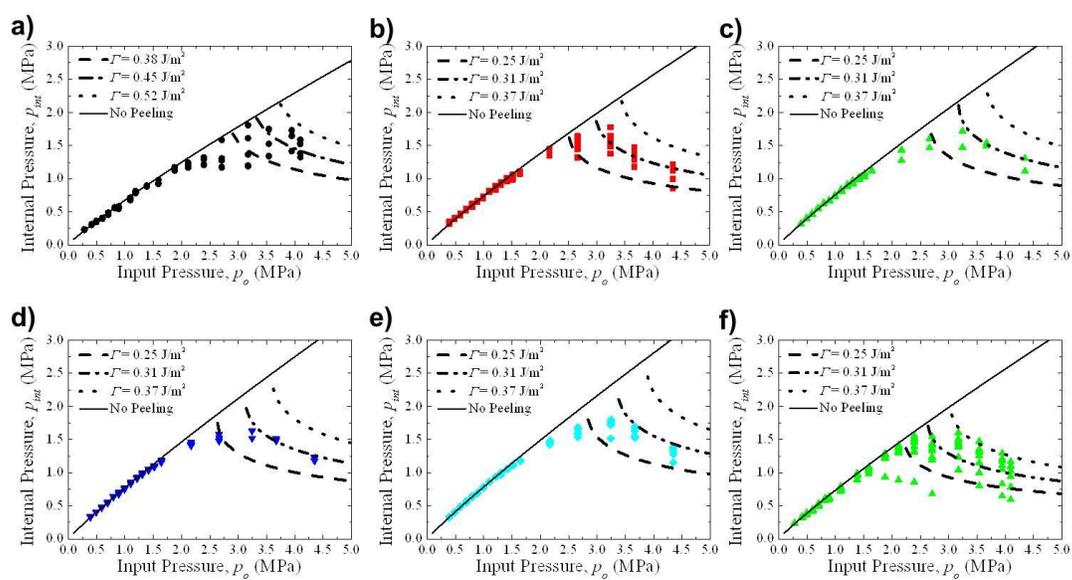

**Figure S4**

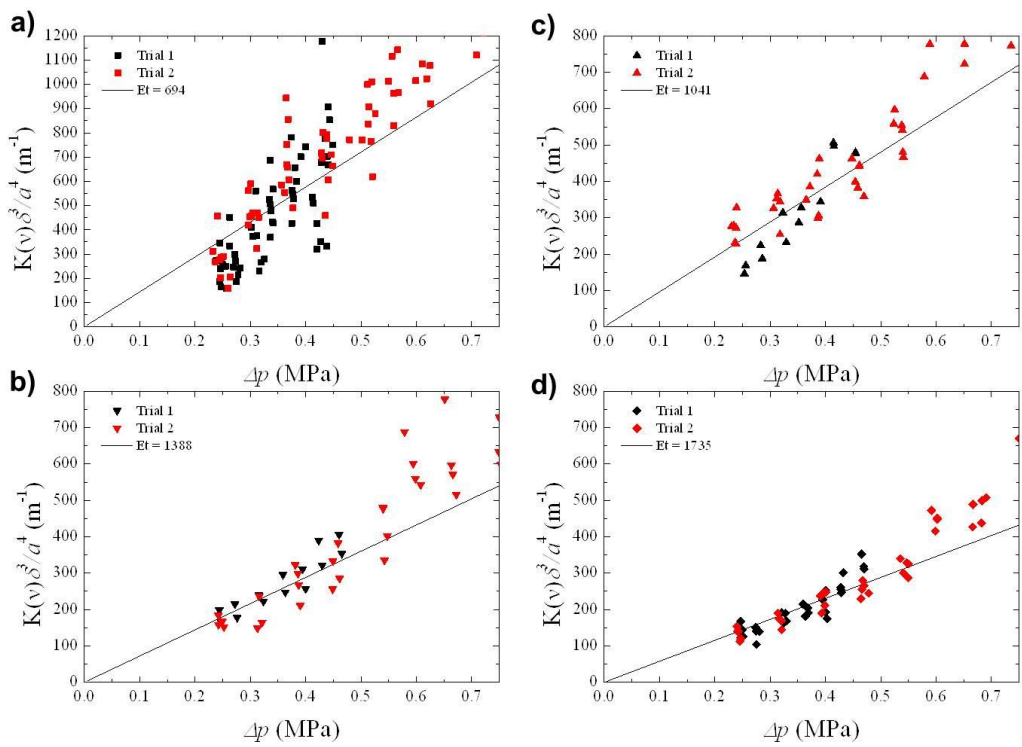

**Figure S5**

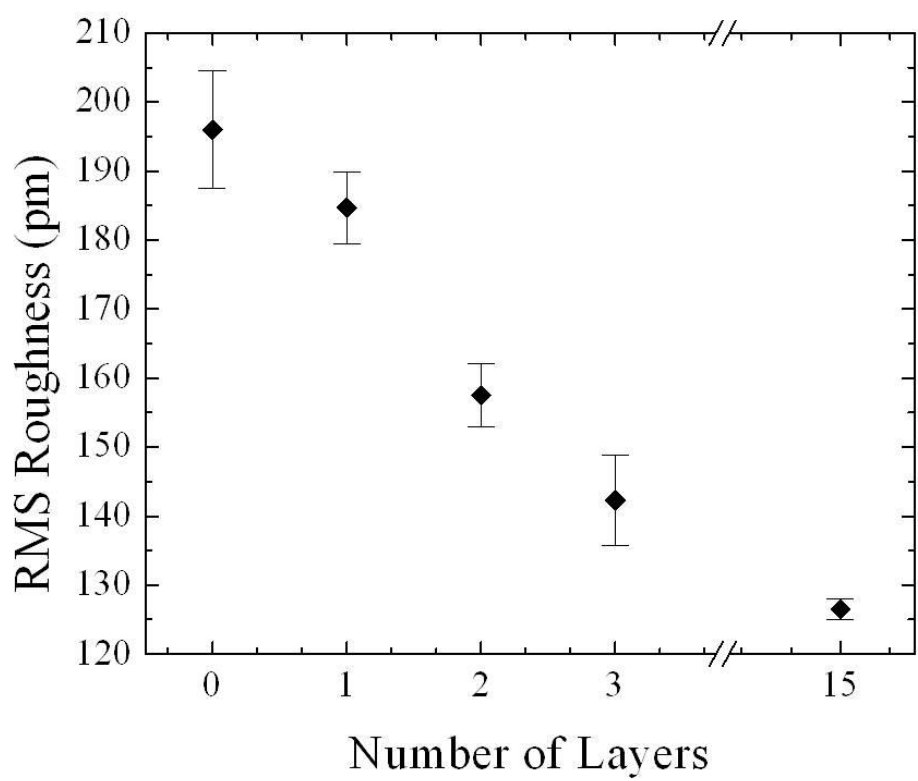

**Figure S6**